%% file: main.tex
\documentclass{article}
\usepackage[utf8]{inputenc}
\usepackage[T1]{fontenc}
\usepackage{hyperref}

\input{style_bibtex}
\usepackage{amsmath,amssymb,amsfonts}
\usepackage{algorithmic}
\usepackage{graphicx}
\usepackage{textcomp}
\usepackage{bmpsize}
\usepackage{xcolor}
\usepackage{lipsum}
\usepackage{microtype}
\usepackage[eandd,preprint,nonatbib]{neurips_2026}
\usepackage{adjustbox}
\usepackage{url}
\usepackage{multirow}
\usepackage{booktabs}
\usepackage{amsfonts}
\usepackage{subfigure}
\usepackage{todonotes}
\usepackage{subcaption}
\usepackage{tabularx}
\usepackage{booktabs}
\usepackage{pifont}
\usepackage{amsmath}
\usepackage{cleveref}
\usepackage{float}
\newcommand{\yes}{{\Large$\bullet$}}
\newcommand{\no}{{\Large$\circ$}}

\newcommand{\assemblage}{\textsc{Assemblage}}
\newcommand{\dataset}{{\textsc{Assemblage-DeepHistory}}}

\usepackage{listings}
\usepackage{xcolor}

\lstdefinestyle{promptstyle}{
    basicstyle=\ttfamily\footnotesize,  %
    breaklines=true,
    breakatwhitespace=true,
    columns=fullflexible,
    frame=single,
    framesep=4pt,
    backgroundcolor=\color{gray!5},
    xleftmargin=4pt,
    xrightmargin=4pt,
    aboveskip=6pt,
    belowskip=6pt,
}

\title{\dataset{}: A Cross-Build Binary Dataset with Temporal Coverage}

\author{%
  Chang Liu\textsuperscript{1} \quad
  Noah Fleischmann\textsuperscript{2} \quad
  Nicolò Altamura\textsuperscript{3} \\
  \textbf{Edward Raff}\textsuperscript{4} \quad
  \textbf{James Holt}\textsuperscript{4} \quad
  \textbf{Kristopher Micinski}\textsuperscript{1} \\[0.6ex]
  \textsuperscript{1}Syracuse University, Syracuse, USA \\
  \textsuperscript{2}Booz Allen Hamilton, McLean, USA \\
  \textsuperscript{3}Independent Researcher, Italy \\
  \textsuperscript{4}CrowdStrike, Austin, USA \\[0.6ex]
  \texttt{\{cliu57,\,kkmicins\}@syr.edu} \quad
  \texttt{Fleischmann\_Noah@bah.com} \\
  \texttt{altamura@nicolo.dev} \quad
  \texttt{\{edward.raff,\,james.holt1\}@crowdstrike.com}
}

\begin{document}

\maketitle

\begin{abstract}

Existing binary corpora typically capture only one or two axes of binary variation: they either provide cross-compiler builds without a temporal axis, or CVE labels for single-build binaries. None combine cross-build diversity, cross-version history, and CVE labels into a queryable structure. We present \dataset{}\footnote{Dataset: \url{https://huggingface.co/datasets/changliu8541/assemblage-deephistory}. Evaluation code: \url{https://github.com/Assemblage-Dataset/DeepHistory-Eval}. Build pipeline (integrated in Assemblage): \url{https://github.com/Assemblage-Dataset/Assemblage}.%
}, which consolidates these dimensions into a unified framework where every binary's compilation context, source code, vulnerable functions, and package version are stored as first-class metadata.

\dataset{} comprises 73,610 binaries spanning 248 open-source projects, compiled across GCC, Clang, and MSVC at multiple optimization levels on Linux and Windows, with multi-year historical builds. Each binary is indexed in a database that links it to its source code, functions, debug info, variant builds, historical versions, and vulnerable functions. Three analyses demonstrate this structure's value: (1) a three-stage LLM benchmark (recognition, strategy-guided detection, and cross-build transfer) to test whether LLMs reason about binary vulnerabilities or pattern-match on build-specific artifacts; (2) a comparison of MalConv embeddings, jTrans function embeddings, and TLSH fuzzy hashes quantifying how same-package versions cluster in each space; and (3) a Bayesian regression decomposing binary similarity into contributions from temporal distance, file changes, and commits.
\end{abstract}

\section{Introduction}\label{sec:introduction}

Binary analysis research is consistently constrained by a structural gap in available datasets: real-world software \emph{evolves} over time. A project is compiled across various compilers, settings, architectures, and ABIs; additionally, code is patched and refactored across updates. Existing corpora capture only fragments of this life cycle. Cross-build datasets~\cite{binkit,jtrans,sigmadiff} vary compiler settings but remain tied to a fixed version, while multi-version efforts~\cite{assemblage,GHTorent,Console2023BinBenchAB,Zuo2024BinSimDBBD} sample many projects but typically retain only a single snapshot of each. CVE-labeled corpora generally associate each vulnerability  with a specific build configuration~\cite{binpool,hussain2025vulbinllmllmpoweredvulnerabilitydetection}. No existing resource combines cross-build diversity, multi-year version coverage, and CVE labels in a structure where compilation context, source origin, and release history are first-class, queryable metadata.

To close this gap, we present \dataset{}, comprising 73,610 binaries (42,188 Windows PE, 31,422 Linux ELF) from 248 open-source C/C++ projects compiled across GCC, Clang, and MSVC at multiple optimization levels. Projects average 5.2 distinct versions and 297 binaries; 140 projects in our corpus span more than two years of development. Every binary is associated via a database schema with its source code, internal functions, sibling builds, subsequent versions, and known vulnerabilities.

Our dataset has three axes: \textbf{(a)} cross-build compilation, \textbf{(b)} multi-year version history, and \textbf{(c)} CVE labels. In combination, \textbf{(a)}--\textbf{(c)} enable us to do three new analyses unavailable on previous, single-axis corpora. First, we conduct several LLM reasoning benchmarks over binaries: recognition, strategy-guided detection, and cross-build transfer. The transfer stage replays the strategy authored on a reference build against compilation variants of the identical vulnerable code, separating semantic reasoning from superficial pattern matching~\cite{ullah2024llmsreliablyidentifyreason,lu2026evaluatingenhancingvulnerabilityreasoning}. Second, leveraging the multi-version axis, we compare MalConv~\cite{malconvGCT} embeddings, jTrans~\cite{jtrans} embeddings, and TLSH~\cite{oliver2013tlsh} fuzzy hashes as signals for clustering same-package binaries. Third, a hierarchical Bayesian regression decomposes binary similarity into contributions from temporal distance, file changes, and commit frequency.

In summary, our primary contributions are as follows:
\begin{itemize}
    \item \textbf{A Multi-Axis Binary Dataset:} We release \dataset{}, a cross-platform corpus of 73,610 binaries spanning 248 open-source C/C++ projects. The dataset is indexed in a queryable database that links every binary to its source code, functions, sibling builds, version history, and 329 CVE-labeled vulnerabilities.
    \item \textbf{A Cross-Build LLM Benchmark:} We construct a three-stage LLM benchmark for binary vulnerability identification. This benchmark leverages the dataset's compilation diversity to explicitly distinguish genuine semantic reasoning from build-specific pattern matching.
    \item \textbf{Similarity and Evolution Analyses:} We characterize the dataset's similarity structure through two complementary lenses: a comparison of MalConv embeddings, jTrans embeddings, and TLSH fuzzy hashes against a hierarchical Bayesian regression that decomposes cross-version binary similarity into temporal, structural, and commit-driven components.
\end{itemize}

\section{Related Work}

\begin{table*}[t]
\centering
\small
\setlength{\tabcolsep}{4pt}
\renewcommand{\arraystretch}{1.1}
\resizebox{\textwidth}{!}{%
\begin{tabular}{lrrrccccc}
\toprule
\textbf{Dataset}
  & \textbf{\begin{tabular}[c]{@{}r@{}}Binaries\\ (\#)\end{tabular}}
  & \textbf{\begin{tabular}[c]{@{}r@{}}Functions\\ (\#, K)\end{tabular}}
  & \textbf{\begin{tabular}[c]{@{}r@{}}Projects\\ (\#)\end{tabular}}
  & \textbf{\begin{tabular}[c]{@{}c@{}}Cross\\ Platform\end{tabular}}
  & \textbf{\begin{tabular}[c]{@{}c@{}}Cross\\ Compiler\end{tabular}}
  & \textbf{\begin{tabular}[c]{@{}c@{}}CVE\\ Included\end{tabular}}
  & \textbf{\begin{tabular}[c]{@{}c@{}}Source\\ Code\end{tabular}}
  & \textbf{\begin{tabular}[c]{@{}c@{}}Temporal\\ Coverage\end{tabular}} \\
\midrule
BinKit~\cite{binkit}                                       & 243,128     & 75,231     & 51        & \no  & \yes & \no  & \no  & \no  \\
BinaryCorp-26M~\cite{jtrans}                               & 48,130      & 25,877     & 9,819   & \no  & \no  & \no  & \no  & \no  \\
BinBench~\cite{Console2023BinBenchAB}                      & 1,127,479 & 4,408      & U         & \no  & \no  & \no  & \no  & \no  \\
LLM4Decompile~\cite{tan2024llm4decompile}                  & U             & U            & 164       & \no  & \no  & \no  & \yes  & \no  \\
$\alpha$diff~\cite{alphadiff}                              & 66,823      & 2,489            & 926         & \no  & \no & \no  & \no  & \yes \\
\textsc{Assemblage}~\cite{assemblage}                      & 1,536,171 & 783,694    & 220,792 & \yes & \no  & \no  & \yes & \no  \\
REALTYPE~\cite{dramko2025idiomsneuraldecompilationjoint}   & U             & 157    & U         & \no  & \no  & \no  & \no  & \no  \\
Decompile-Bench~\cite{tan2025decompilebench}               & 85,000             & 100,000      & 3,961         & \no  & \no  & \no  & \yes & \no  \\
BinPool~\cite{binpool}                                     & 6,144           & 7            & 162       & \no  & \no  & \yes & \no  & \no  \\
\midrule
\textbf{\textsc{\dataset{}}}
& \textbf{73,610}
& \textbf{441,858}
& \textbf{248}
& \textbf{\yes}
& \textbf{\yes}
& \textbf{\yes}
& \textbf{\yes}
& \textbf{\yes} \\
\bottomrule
\end{tabular}%
}
\caption{Comparison of binary datasets in reverse engineering, decompilation, and vulnerability research. \yes\ indicates the feature is present, \no\ indicates absent. U: undisclosed or unknown.}
\label{tab:dataset-comparison}
\end{table*}

\paragraph{Binary Datasets for Reverse Engineering}

The advancement of binary analysis research is fundamentally dependent on large-scale corpora of compiled programs. Therefore, substantial work has focused on constructing these datasets by crawling open-source repositories and employing automated build drivers (such as Make and CMake) to compile the source code at scale~\cite{NIPS2007_a532400e,GHTorent,10.1145/3379597.3387481,Shao_2020,ghcc,Shariati2015UbuntuOI,280046}. This methodology yields extensive collections of ELF binaries across different compiler versions and optimization levels, enabling downstream applications like binary similarity, function name recovery, and decompilation ~\cite{liu2026supersetdecompilation,7546500,294518,zou2025dliftimprovingllmbaseddecompiler,8077799,tan2025decompilebench}. Nevertheless, these automated compilation pipelines remain overwhelmingly restricted to Linux. While standardized build systems and package managers render headless compilation tractable in Linux environments, equivalent uniformity does not exist for Windows.

To address this limitation, \textsc{Assemblage}~\cite{assemblage} expanded corpus collection to Windows by compiling hundreds of thousands of projects across both operating systems, supplementing GitHub-derived source code with curated vcpkg~\cite{vcpkg} packages. Other corpora target highly specific reverse-engineering domains. For instance, in the realm of similarity research, BinKit~\cite{binkit} provides binaries compiled across various optimization variants, while $\alpha$Diff~\cite{alphadiff} contributes a massive dataset for cross-version similarity detection, the data only covers the temporal axis without cross-compiler/optimization/platform variants. Similarly, LLM4Decompile~\cite{tan2024llm4decompile}, NOVA~\cite{jiang2025novagenerativelanguagemodels}, and IDIOMS (REALTYPE dataset)~\cite{dramko2025idiomsneuraldecompilationjoint} focus on neural decompilation, while EMBER~\cite{Anderson2018EMBERAO,EMBER2024,Corlatescu2023EMBERSimAL} and BODMAS-style malware corpora~\cite{joyce22,Console2023BinBenchAB,NEURIPS2024_2663c994} support classification research. While these datasets significantly advance the field, they typically capture only a single snapshot of each project. Even datasets that include multiple variants fail to capture combined cross-build and cross-version dynamics with comprehensive contextual information. This leaves a critical gap in understanding how compilation choices and version drift interact over years of software evolution.

\paragraph{Cross-Build and Multi-Version Binary Analysis}
Motivated by critical tasks such as patch identification, vulnerability propagation, and malware variant detection, binary similarity research has historically focused on cross-version and cross-build comparisons. Classical approaches rely on graph-matching tools~\cite{flake2004structural,diaphora,feng2016scalable,ktrans,xiaomvp2020,binhunt} to compare control-flow graphs across compilations. The field later shifted toward neural methods that replace graph isomorphism with learned embeddings, evolving from early architectures (Gemini, a graph-embedding similarity model~\cite{geminiccs}, $\alpha$Diff~\cite{alphadiff}, SAFE~\cite{safeembed}, DeepBinDiff~\cite{DBLP:conf/ndss/DuanLWY20}) to transformer-based models with extensive pretraining (PalmTree~\cite{palmtree}, BinProv~\cite{He2022BinProvBC}, jTrans~\cite{jtrans}, SigmaDiff~\cite{sigmadiff}). However, while these methods are evaluated on cross-build pairs, their underlying training corpora remain inherently planar: they either vary compiler settings at a single snapshot or span a minimal version axis (e.g., adjacent releases) under a fixed build configuration. Recent library identification and version fingerprinting models~\cite{duan2017osspolice,yuan2019b2sfinder,jiang2024binaryai,zou2025bincofer,dong2024libvdiff} inherit this identical flat structure. \dataset{} closes this planar gap by filling out the third axis of temporal coverage, seamlessly pairing cross-build diversity (compilers, optimization levels, operating systems) with deep, multi-year histories that span more than two years for 140 of our 248 packages.

\paragraph{Cross-Build Vulnerability Binary Datasets for LLM Evaluation}
LLMs are widely suspected of succeeding on benchmarks through memorization or surface pattern matching rather than reasoning over program semantics~\cite{liu2024lost,hsieh2024ruler,wei2025equibench,zhou2024donttrustverify,yang2026eligibilityllmscounterfactualreasoning,riddell-etal-2024-quantifying}. To address this, equivalence-style benchmarks such as EquiBench~\cite{wei2025equibench} hold program behavior fixed while varying syntactic form. At the source level, this philosophy supports a mature ecosystem: SecVulEval~\cite{ahmed2025secvulevalbenchmarkingllmsrealworld}, CVE-Bench~\cite{wang-etal-2025-cve}, SafeGenBench~\cite{li2025safegenbenchbenchmarkframeworksecurity}, and SEC-bench~\cite{lee2025secbenchautomatedbenchmarkingllm} all draw on richly annotated CVE corpora that track vulnerability-fixing commits across versions~\cite{bigvul,cvefixes,ni2024megavul,ding2024primevul,VulZoo}. At the binary level, however, the landscape is far sparser. BinPool~\cite{binpool} labels 603 Linux CVEs on fixed compilations, and Vul-BinLLM~\cite{hussain2025vulbinllmllmpoweredvulnerabilitydetection} relies on synthetic Juliet cases. Because both tie each vulnerability to a single build, they leave no way to separate vulnerability semantics from build-specific artifacts. \dataset{} extends the equivalence paradigm to compiled code: the underlying flaw retains its semantic identity while the compiler, optimization level, and OS may vary, ensuring that successful detection strategy must generalize across builds.

\section{Dataset Details}\label{sec:methods}

\begin{table}[t]
\centering
\small
\setlength{\tabcolsep}{6pt}
\renewcommand{\arraystretch}{1.1}
\begin{tabular}{lrrrrrr}
\toprule
& \textbf{Projects} & \textbf{Binaries} & \textbf{PE} & \textbf{ELF} & \textbf{Avg ver.} & \textbf{Bins/Proj} \\
\midrule
\multicolumn{7}{l}{\textit{By temporal coverage}} \\
0--1 year     & 77  & 13,529  &  5,303 &  8,226 &  2.2 & 176 \\
1--2 years    & 31  & 10,471  &  7,919 &  2,552 &  5.8 & 338 \\
2--5 years    & 118 & 42,454  & 25,473 & 16,981 &  6.7 & 360 \\
5--10 years   & 22  &  7,156  &  3,493 &  3,663 &  6.9 & 325 \\
\midrule
\multicolumn{7}{l}{\textit{By distinct versions}} \\
1      & 46  &  2,210  &  1,587 &    623 &  1.0 &  48 \\
2--5   & 105 & 22,216  & 11,264 & 10,952 &  3.9 & 212 \\
6--10  & 83  & 40,664  & 25,260 & 15,404 &  7.6 & 490 \\
11--20 & 12  &  7,191  &  3,007 &  4,184 & 12.8 & 599 \\
20+    & 2   &  1,329  &  1,070 &    259 & 27.0 & 664 \\
\midrule
\textbf{Total} & \textbf{248} & \textbf{73,610} & \textbf{42,188} & \textbf{31,422} & \textbf{5.2} & \textbf{297} \\
\bottomrule
\end{tabular}
\caption{Coverage of \dataset{} (library-only binaries) grouped by version
span (top) and number of distinct versions per project (bottom).
\emph{Avg ver.}\ is the mean number of distinct versions per project in the
binaries; \emph{Bins/Proj} is the mean binaries per project.}
\label{tab:coverage}
\end{table}

\subsection{Dataset Construction}
Our corpus targets foundational C/C++ libraries characterized by well-documented multi-version histories. The selection criteria included popularity, breadth of integration, and sustained maintenance activity. To ensure our dataset remains complementary, we deliberately excluded vcpkg~\cite{vcpkg} repositories previously compiled by \assemblage{}. To handle historical compilations, we employ two distinct strategies: for Windows, we utilize Conan's standardized recipes to automate dependency resolution and compilation; for Linux, we build historical versions directly from their upstream GitHub repositories.

By compiling across both operating systems, the \dataset{} pipeline systematically generates Windows PE binaries (preserving MSVC PDB debug files) alongside Linux ELF binaries (retaining GCC/Clang DWARF sections). This dual-platform approach resolves two major gaps in prior Linux-centric corpora~\cite{assemblage,ghcc}: it reconstructs historical Windows versions that lack centralized package snapshots (e.g., \texttt{apt} or \texttt{yum}), and natively captures the Windows ecosystem's heavy reliance on bundled, dynamically linked libraries.

\subsection{Dataset Composition}

\dataset{} comprises 73,610 compiled library binaries (42,188 Windows PE and 31,422 Linux ELF) sourced from 248 software packages. The corpus contains approximately 442 million functions, all indexed within a central SQLite database that preserves the mapping between source repositories, function metadata, and build configurations. Each binary is compiled across multiple configurations: Windows binaries are built with MSVC under \textit{Debug} and \textit{RelWithDebInfo} profiles (generating external PDB files), while Linux binaries utilize GCC and Clang with optimizations from \texttt{-O0} to \texttt{-O3} (retaining in-place DWARF sections). This debug information provides high-fidelity ground truth—including function boundaries, symbol names, and source-line mappings—essential for supervised function-level analysis.

As detailed in Table~\ref{tab:coverage}, the dataset offers significant temporal depth: 140 packages span more than two years of development, with an average of 5.2 versions per project. The symbol information recovered from PDB and DWARF headers demonstrates remarkable cross-version stability; in prominent packages, tens of thousands of functions persist across numerous version pairs. This longitudinal consistency enables rigorous study of function evolution and cross-version matching without the noise introduced by heuristic boundary detection.

\section{Evaluations}\label{sec:eval}

We demonstrate \dataset{}'s research value through three analyses spanning different aspects of the dataset's structure: Section~\ref{sec:eval-recognition} introduces a three-stage LLM benchmark that uses the cross-build/cross version axis to test vulnerability reasoning; Section~\ref{sec:similarity} compares three off-the-shelf binary representations on the multi-version axis; and Section~\ref{sec:bayesian} decomposes binary similarity into temporal, structural, and activity components.

\subsection{LLM Vulnerability Understanding}
\label{sec:llmvul}

\begin{figure}
    \centering
    \includegraphics[width=1\linewidth]{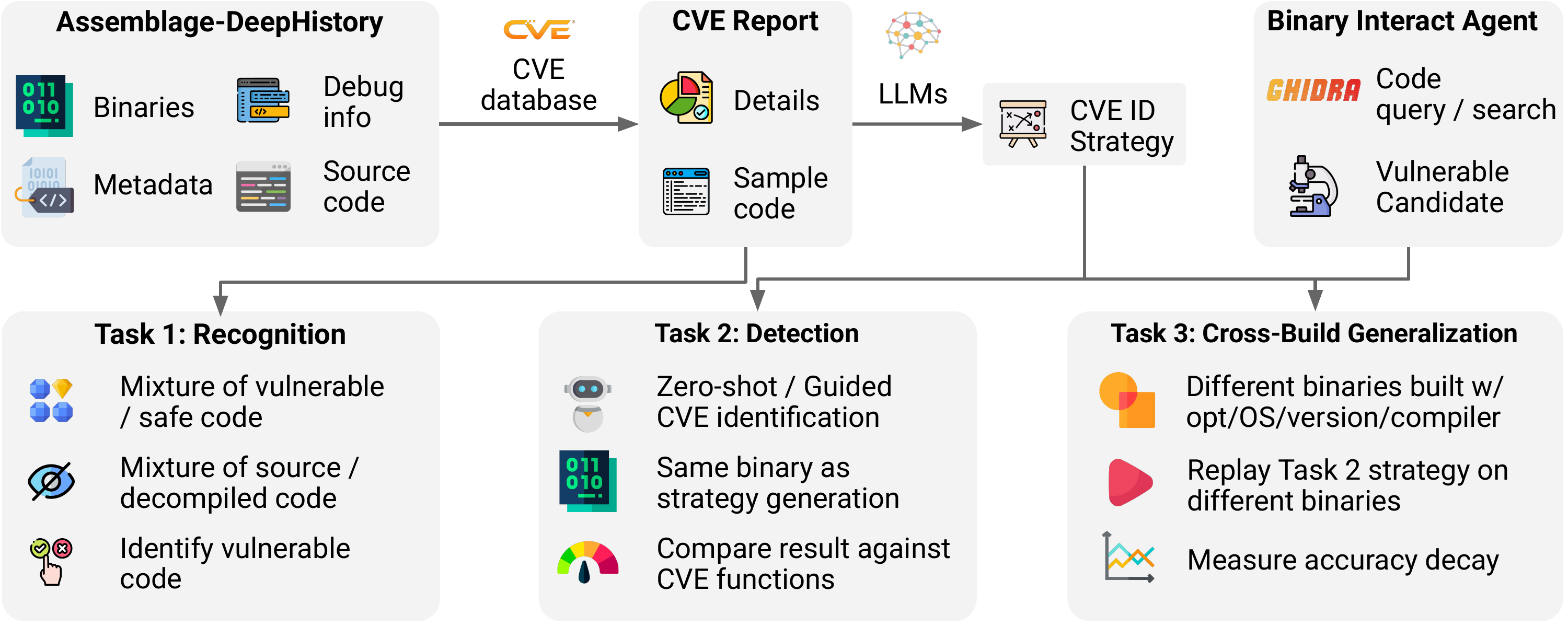}
    \caption{Three-Stage CVE Evaluation Design}
    \label{fig:cvebenchdesign}
\end{figure}

\paragraph*{Experiment Setup}
\label{part:specs}

Our benchmark contains 329 CVEs across 55 packages from \dataset{}, and the design is illustrated in Figure~\ref{fig:cvebenchdesign}. CVEs are fetched by package name from the National Vulnerability Database, the CVE Program, and the GitHub Advisory Database~\cite{cve,ghsa,nvd} then filtered to the ones that affected versions intersect with compiled binary versions in the dataset. Then, we use the CVE description and patch diff to identify affected functions and we inspect about 25\% of the resulting records to verify that each CVE matches the correct library and version in our dataset (manually inspected CVE IDs available in appendix). For each CVE, we chose a reference binary with lowest optimization and grouped other affected binaries to one of five Diff categories: Optimization, Compiler, OS, Version and All (per-category counts in the appendix).

Experiments were conducted on a HTCondor managed cluster, and each environment was running Ubuntu 20.04.5 LTS, equipped with an AMD EPYC 9845 160-core CPU, an NVIDIA L40S GPU, and 502 GB of memory. Qwen-3.6~\cite{qwen3} (Qwen/Qwen3.6-35B-A3B) was served via SGLang with FP8 quantization, and Gemma 4\cite{gemma4} (gemma4:26b) was served via Ollama with Q4\_K\_M quantization ( Gemma 4 had compatibility issues with SGLang at the time of the experiments). Both models were configured with a 256K-token context window and temperature=0. Ghidra~\cite{ghidra} version 12.0.3 was used in the evaluations. Evaluations in Section~\ref{sec:llmvul} consumed roughly 1800 L40S GPU-hours, in 8 days of queued cluster time. The remaining models used in our evaluations, Opus~4.7, GPT-5.4, and Gemini~3.1~Pro Preview (will be later referenced as Opus, GPT, and Gemini) were accessed with their latest version at the time of experiment.

\paragraph*{Binary CVE Recognition} \label{sec:eval-recognition} 
In the first and simplest task, the model selects the vulnerable function from $K=5$ candidates from a single binary. The candidates are made from the vulnerable function and four distraction choices, and the four are chosen sorted by CodeBLEU~\cite{ren2020codebleu} similarity, and shuffled before prompting. $K=5$ gives a small 20\% random baseline while keeping the large CVE prompts within local agents' $256$K context, avoiding long-context degradation~\cite{liu2024lost,hsieh2024ruler}. We evaluate each model under two independent factors: code representation (Ghidra-decompiled vs.\ original source code) and vulnerability context (zero-shot vs.\ with the CVE description and patch diff).

Table~\ref{tab:recognition} reports accuracy across five models. Note that Gemini 3.1 Pro fails to return a final answer on 1.8\% even after several attempts, and these are marked as incorrect. Every model loses accuracy moving from source to binary (-0.05 to -0.20, smaller for frontier models) and gains substantially with the CVE description (+0.46 to +0.60). With description, the three frontier models reach high accuracy on both source and binary inputs; however on zero-shot tasks, all five model answers are near baseline, indicating that recognition without grounding CVE context text is close to random even for the strongest models.

\label{sec:eval-recognition}

\begin{table}[t]
\centering
\small
\setlength{\tabcolsep}{6pt}
\renewcommand{\arraystretch}{1.15}
\begin{tabular}{lccccc}
\toprule
\textbf{Setting} & \textbf{Qwen} & \textbf{Gemma} & \textbf{Opus} & \textbf{GPT} & \textbf{Gemini} \\
\midrule
Source, zero-shot       & 0.27 & 0.26 & 0.26 & 0.27 & 0.27 \\
Source, with-desc       & 0.85 & 0.79 & 0.88 & 0.87 & 0.88 \\
Binary, zero-shot       & 0.16 & 0.13 & 0.20 & 0.18 & 0.24 \\
Binary, with-desc       & 0.61 & 0.52 & 0.78 & 0.75 & 0.82 \\
\midrule
$\Delta$ decompilation  & $-0.18$ & $-0.20$ & $-0.08$ & $-0.10$ & $-0.05$ \\
$\Delta$ description    & $+0.51$ & $+0.46$ & $+0.60$ & $+0.59$ & $+0.60$ \\
\bottomrule
\end{tabular}
\caption{CVE recognition accuracy across models on CVE recognition task. $\Delta$ decompilation is the mean accuracy drop from source to binary, averaged over the zero-shot and with-desc settings; $\Delta$ description is the mean gain from providing the CVE description, averaged over source and binary inputs. Random baseline: 0.20.}
\label{tab:recognition}
\end{table}

\paragraph*{Guided CVE Detection} \label{sec:eval-hunting} 
In the second task, an agent aims to locate the vulnerable function in a stripped binary by exploring it through \texttt{BinaryAPI}, a read-only $14$-method interface we built over Ghidra-based analysis.

\texttt{BinaryAPI} exposes three code representations per function (decompiled C, annotated post-SSA p-code, and raw disassembly), bidirectional call-graph navigation, control-flow graphs and regex search across each representation. Each \texttt{BinaryAPI} call is a deterministic lookup against cached analysis (regex call executes at runtime), and calls execute identically across agents. The interface does not allow arbitrary code execution or debug symbols exposure. We initially conducted 100 pilot runs to set time and turn limits that give agents enough room to work without hanging the GPU cluster; these runs used a 30-minute, 1,000-turn cap and are not included in the final results (running-time and turn count statistics in the appendix). For the full evaluation we loosened these limits to a 1-hour wall-clock budget with no turn cap, so that agents are not cut off early. The agent terminates when the model commits to up to five candidates or returns 10 consecutive empty messages. A task scores Hit@$k$ if any ground-truth function appears in the top-$k$ answers; for multi-function CVEs, the best-ranked match is counted. We evaluate execution agents Gemma~4 and Qwen~3.6 in two settings: solo (locate vulnerable function given no context) and strategy-guided (strategies from five generator LLMs based on CVE, patch). Prompts are provided in the appendix. This way, we measure the generator LLMs' understanding of vulnerability apart from its tool-calling capabilities, and the solo results provide a baseline to avoid noise introduced by agent execution.

Table~\ref{tab:cve-hunting} reports solo Hit@$k$ alongside strategy-guided Hit and uplifts. The Qwen agent benefits substantially more than the Gemma agent, reaching Hit@$1$ of 0.83 under Opus-authored strategies versus 0.29 for the Gemma agent on the same strategies. Opus produces the most effective strategies for the Qwen agent, leading on both Hit@$1$ and Hit@$5$, while Qwen-authored strategies lead for the Gemma agent. The Hit@$1$ spread across the five generators is also wider for the Qwen agent ($0.38$--$0.83$, std $\approx 0.16$) than for the Gemma agent ($0.15$--$0.32$, std $\approx 0.06$). This pattern suggests that tool-calling capability limits the effect of strategy quality: only when an agent can reliably execute the strategy, the strategic differences can affect the outcomes. The Qwen agent's wider spread therefore reflects its stronger understanding and execution capability on this task.

\begin{table}[t]
\centering
\small
\setlength{\tabcolsep}{4pt}
\renewcommand{\arraystretch}{1.2}
\begin{tabular}{llcccccc}
\toprule
& & \textbf{Solo} & \multicolumn{5}{c}{\textbf{Strategy-guided (Hit / $\Delta$)}} \\
\cmidrule(lr){4-8}
\textbf{Agent} & \textbf{Metric} & \textbf{(baseline)} & \textbf{Gemma} & \textbf{Qwen} & \textbf{Opus} & \textbf{GPT} & \textbf{Gemini} \\
\midrule
\multirow{2}{*}{Gemma-agent}
& Hit@1 & 0.00 & 0.15 / +0.15 & 0.32 / +0.32 & 0.29 / +0.29 & 0.24 / +0.24 & 0.29 / +0.29 \\
& Hit@5 & 0.02 & 0.21 / +0.19 & 0.34 / +0.32 & 0.31 / +0.29 & 0.27 / +0.25 & 0.31 / +0.29 \\
\midrule
\multirow{2}{*}{Qwen-agent}
& Hit@1 & 0.03 & 0.38 / +0.35 & 0.72 / +0.69 & 0.83 / +0.80 & 0.75 / +0.72 & 0.78 / +0.75 \\
& Hit@5 & 0.12 & 0.53 / +0.41 & 0.82 / +0.70 & 0.86 / +0.74 & 0.81 / +0.69 & 0.85 / +0.73 \\
\bottomrule
\end{tabular}
\caption{CVE locating performance on reference binaries (Eval~2). Each cell averages over agent results on 329 CVEs; Solo is the agent's no-strategy baseline; $\Delta$ is the uplift over the Solo baseline. Values are Hit@$k$ accuracy in $[0, 1]$.}
\label{tab:cve-hunting}
\end{table}

\paragraph*{CVE Cross-Build Generalization} \label{sec:eval-transfer}
The third task tests the generalization and understanding of the LLMs on vulnerability. We first verify that the build axes produce genuinely different decompiled code. On decompiled code pairs spanning 100 CVEs (259 Linux ELF and 194 Windows PE binaries; full table in appendix), CodeBLEU~\cite{ren2020codebleu} drops with build divergence: 0.53--0.69 within-compiler cross-config, 0.45 cross-compiler within Linux, and 0.29 cross-OS.

We execute the Eval~2 strategy-guided agents on every cross-build variant (Opt, Compiler, OS, Version, All), reusing the Eval~2 infrastructure. We report only the Qwen agent; Gemma's reference-build accuracy is too low across all generator LLMs for cross-build decay to be measurable, and the results are deferred to the appendix.

\begin{figure}[t]
  \centering
  \includegraphics[width=\linewidth]{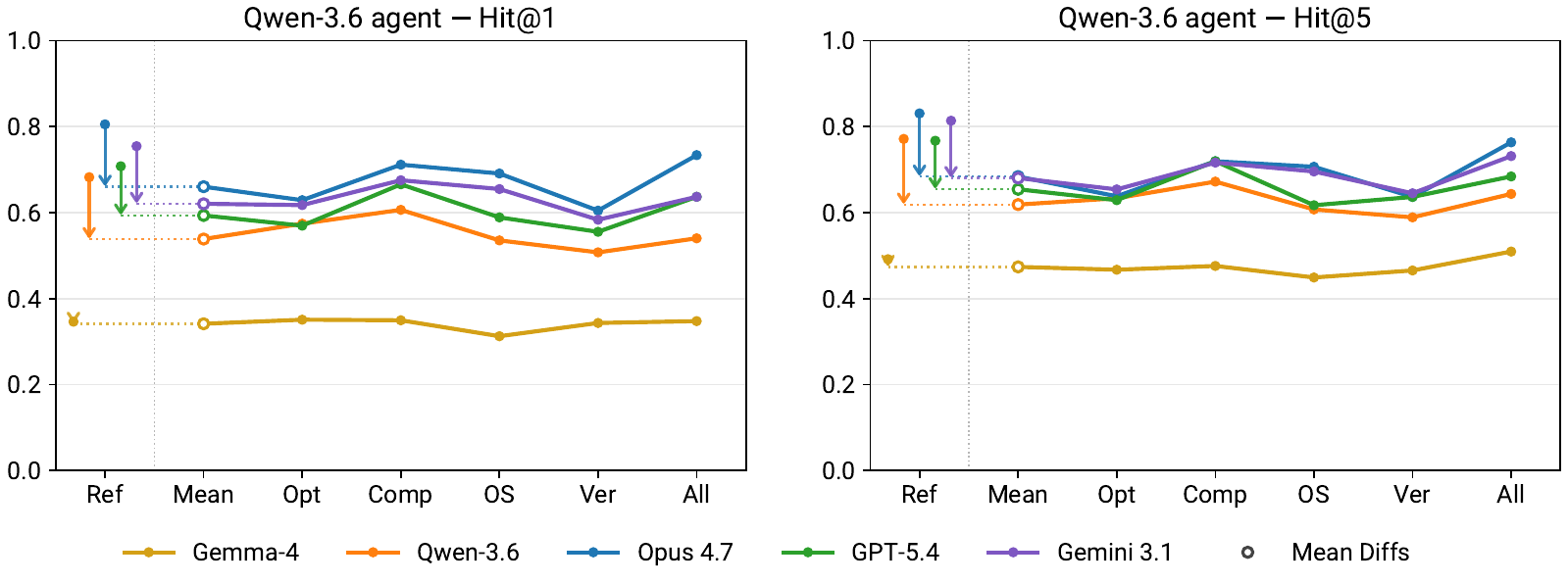}
    \caption{Cross-build transfer based on Qwen-3.6 agent. Each panel plots Hit@$1$ (left) or Hit@$5$ (right) across the Reference binary and the five Diff buckets (Opt, Compiler, OS, Version, All), with one line per strategy author LLM. The leftmost ``Mean Diffs'' marker pools the five Diff buckets. Gemma agent results are deferred to the appendix.}
  \label{fig:eval3-transfer}
\end{figure}

Figure~\ref{fig:eval3-transfer} plots the Qwen agent's Hit@$1$ and Hit@$5$ across the Reference binary, five Diff buckets, and their pooled Mean. A strategy that captures vulnerability semantics should remain effective even when the binary diverges from the build it was authored against. We observe that all four strong-generator strategies accuracy decrease from Reference to the Diff mean by similar absolute amounts (Hit@$1$ drops of $0.12$--$0.15$). In addition, Diff~OS (CodeBLEU $0.29$) is not the empirically hardest axis, and Diff~Version is the lowest Hit@$1$ point for all four strong generators, indicating a shared blind spot across these strategies on software evolution. Hit@$5$ shows a similar ordering at higher values, with Opus and Gemini effectively tied at the Diff mean ($\approx 0.7$).

\subsection{Package-Level Binary Similarity}
\label{sec:similarity}

To test how well off-the-shelf binary representations cluster same-package versions, we compare three signals on \dataset{}: MalConv embeddings~\cite{malconvGCT}, a CNN pretrained on Windows PE for malware classification; jTrans~\cite{jtrans}, a transformer pretrained for binary function similarity; and TLSH fuzzy hashes~\cite{oliver2013tlsh}, a locality-sensitive byte hash. All three are evaluated as out-of-the-box, untuned representations. The per-pair decomposition of similarity into temporal, structural, and activity components is the subject of Section~\ref{sec:bayesian}; here we measure only what each representation provides under arbitrary cross-build pairs. Of \dataset{}'s 248 projects, 52 have only one version or don't come with .dll/.so/.lib and are excluded. For TLSH we follow the original paper's practice~\cite{oliver2013tlsh} and treat distance $\leq 100$ as similar; each package cell reports the fraction of within-package pairs meeting that threshold.

\begin{figure}[t]
    \centering
    \includegraphics[width=\linewidth]{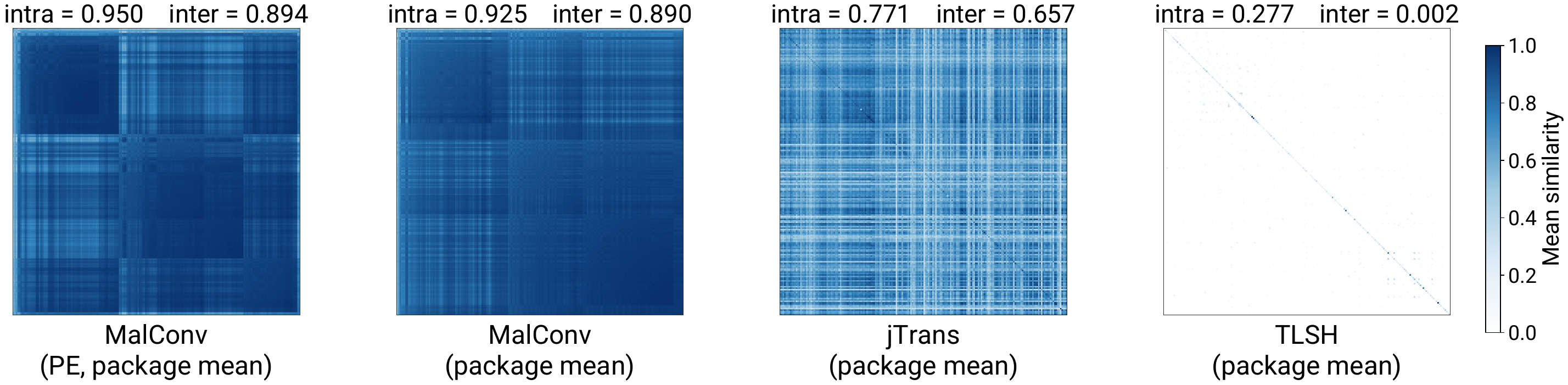}
    \caption{Package-level binary similarity on \dataset{}'s $\geq 2$-version subset (ELF + PE combined). From left to right: MalConv embedding cosine similarity, PE only; MalConv embedding cosine similarity, all packages-mean; jTrans embedding cosine similarity, all packages-mean; TLSH fuzzy-hash similarity, all packages-mean.}
    \label{fig:similarity-comparison}
\end{figure}

Figure~\ref{fig:similarity-comparison} shows package-mean similarity matrices for all three signals. The two MalConv panels (left, middle) show only modest separation between intra- and inter-package pairs on the full corpus (pair-level Cohen's $d = +0.21$), and the gap remains similarly limited on the Windows PE sub-population alone. Because the same pattern holds on the PE binaries that match MalConv's training distribution, the limited separation cannot be attributed to applying PE weights to ELF inputs. Therefore, pretrained MalConv weights do not provide a usable signal for package-level similarity or version-differentiation tasks on this corpus. jTrans (third panel) shows modestly better separation (intra-package mean $0.771$ vs.\ cross-package mean $0.657$), but both means remain high in absolute terms, indicating that a transformer pretrained for binary function similarity also fails to cleanly distinguish same-package from different-package binaries on this corpus.

TLSH, in contrast, yields a sparse similarity space (right panel): intra-package mean $0.277$ versus cross-package mean $0.002$ (with pair-level Cohen's $d = +1.36$). TLSH also recovers genuine cross-package code reuse, for example \texttt{libnghttp2}$\leftrightarrow$\texttt{nghttp3} at $0.345$ captures shared structure between related HTTP/2 and HTTP/3 implementations. TLSH is therefore the most usable of the three off-the-shelf signals on \dataset{}, supporting both within-package coherence and detection of cross-package shared code; the strong intra/inter contrast further confirms that \dataset{} captures meaningful variation across packages.

\subsection{Hierarchical Bayesian Regression}
\label{sec:bayesian}

\dataset{} also allows for investigation of the relationship between code evolution and binary similarity, which can be critical for maintaining databases of known functions that are used for both benign and malicious purposes (e.g., encryption). We perform a Bayesian analysis to demonstrate the avenue of research this enables, comparing MalConv cosine similarity, jTrans cosine similarity, and TLSH transformed similarity as response variables under the same hierarchical model. The model is computed using Markov Chain Monte Carlo~\cite{metropolis1953a} with the NUTS sampler~\cite{hoffman2014a} and implemented with Numpyro~\cite{phan2019composableeffectsflexibleaccelerated}. The $\hat{R}$ scores for all reported global coefficients were $\leq 1.01$, indicating convergence~\cite{b03a7d33-4a4a-354b-bca4-ded21a5d7dd0}. All distributions were Gaussian or half-Gaussian (positive values only) as appropriate, and the target variable was modeled as a Beta regression: for each release pair $i$, the observed similarity $y_i \in (0,1)$ is modeled as $y_i \sim \mathrm{Beta}(\mu_i\kappa, (1-\mu_i)\kappa)$, where $\mathrm{logit}(\mu_i)$ is a linear function of standardized covariates and project-specific coefficients, with concentration $\kappa \in \mathbb{R}^+$. As TLSH produces an unbounded nonnegative distance rather than a similarity score, we transform TLSH distance $d$ into a Beta-regression response using $s = (1+\exp((d-m)/\tau))^{-1}$, where $m$ is the median TLSH distance and $\tau$ is chosen so that the 90th percentile maps to $0.1$ (current values from data: $m = 142.35$, $\tau = 33.44$). We then apply the standard open-interval squeeze so that all responses lie in $(0, 1)$.

\begin{figure}[]
    \centering
    \includegraphics[width=0.8\linewidth]{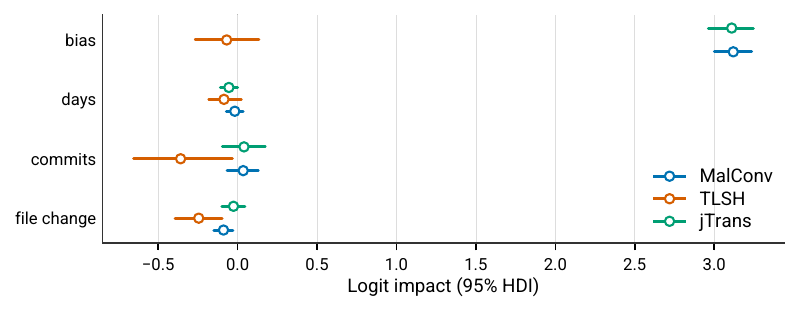}
    \caption{Global coefficient posterior means and 95\% HDIs for MalConv cosine similarity, jTrans cosine similarity, and transformed TLSH similarity. Using three methods of similarity, we can gain better insights into the likely strong factors of code variation.}
    \label{fig:hyperPrior}
\end{figure}

The Bayesian model analysis sought to understand the factors that influence binary similarity across versions. Each feature of interest is modeled with a hyper-prior (the global rate), which is used as the prior for a project-specific version of the covariate. In this way, the Bayesian approach shares information across all projects while also recognizing that the multiple compilations of one project have an intra-source similarity that should not influence other projects~\cite{Gelman2012}.  

We constructed the model using three covariates: the number of days between each pair of releases, the number of commits between them, and the number of changed source files normalized by the total source-file count in the base release. We fit the same hierarchical model separately for MalConv cosine similarity, jTrans cosine similarity, and transformed TLSH similarity. The temporal distribution of releases in \dataset{} is non-uniform: release activity grows steadily, peaking in 2023--2024 with sustained activity throughout. As a result, many version pairs have short temporal separations while fewer span the full range, directly influencing how we interpret the relationship between release timing and binary similarity. Given these temporal characteristics, we first validate that the model appropriately captures the underlying relationships using a posterior predictive check that compares predictions with observed values~\cite{10.1214/06-BA117A}.

Resolving a model that ``fully captures'' the distribution, or determining whether such a model exists, is beyond the scope of this article which introduces the dataset and demonstrates the potential research value. We can examine the hyper-prior's Highest Density Interval (HDI, a Bayesian counterpart to the frequentist confidence interval~\cite{kruschke2014doing}) to determine if a statistically significant relationship exists between the covariates and outcome (file similarity).

The strongest consistent covariate effect is normalized file change, which is negative for both MalConv and TLSH, while the HDI crosses zero for jTrans, with a small negative posterior mean. This indicates that releases touching a larger fraction of the source tree tend to produce less similar binaries, consistent with widespread file changes signaling architectural refactoring or dependency updates rather than incremental feature additions. Commit count differs by metric: its HDI crosses zero for MalConv and jTrans, but is credibly negative for TLSH, where commits carry the largest-magnitude effect of any covariate. This suggests TLSH is more sensitive than MalConv and jTrans to accumulated byte-level change across active development histories. Calendar time has a negative posterior mean for all three metrics. Its HDI is credibly negative for MalConv but crosses zero for TLSH and jTrans, so release age is a stable predictor only for the MalConv embedding similarity, not for TLSH or jTrans after accounting for commits and file change.

The large positive MalConv and jTrans biases reflect the high baseline similarity of these embeddings across versions, consistent with the saturation observed in Section~\ref{sec:similarity}. The TLSH bias is closer to zero after the sigmoid distance transform, reflecting a less saturated similarity scale in which structural divergence is more readily expressed. The results from Figure~\ref{fig:hyperPrior} are again the hyper-prior measuring global rates. A Bayesian model allows us to simultaneously estimate the per-project local rates. Inspecting the per-project offsets from the global prior shows that most projects cluster near the global rate, which is necessary for the linear model to be a reasonable choice, as confirmed by the posterior predictive check. The per-project offsets also exhibit covariate-specific heterogeneity, indicating author or project-specific differences in development cadence and providing insight into factors influencing binary evolution beyond the shared global rate.

\section{Conclusion, Limitations, and Future Work}
\label{sec:conclusion}

\dataset{} unifies cross-build diversity, multi-year version history, and CVE-annotated vulnerability labels in a single queryable corpus, linking each binary to its source, sibling builds, historical versions, and vulnerability metadata. Across our three evaluations, a common lesson emerges: signals that appear stable in isolation---an LLM agent's strategy, a pretrained embedding, or a binary-similarity score---may change once the same source-level semantics are observed across different builds and releases. These results show why these separate axes of variation must be studied jointly: their interaction exposes effects that are obscured by single-axis corpora.

\paragraph{Limitations} CVE coverage concentrates on 55 of the 248 projects, reflecting the uneven distribution of vulnerability disclosure across open source. In addition, \dataset{} targets x86 due to its dominant share and because our build hosts are natively x86; ARM or RISC-V coverage would require cross-compilation or a different host architecture, which is left to future work.

\dataset{}'s vulnerability component is intended for defensive research and carries the dual-use tension common to any CVE-labeled corpus. We mitigate this by including only \emph{publicly disclosed, already-patched} CVEs whose vulnerable functions and patch diffs already appear in the NVD and upstream commits. The source is public and our binaries are built unmodified from upstream releases, thus the corpus adds no offensive capability.

\paragraph{Future Work}

\dataset{} is built from unmodified upstream source; adversarially obfuscated binaries are out of scope despite a rich literature and trend on obfuscation and obfuscation-resilient analysis~\cite{collberg1997taxonomy,junod2015obfuscatorllvm,tigress,schloegel2022loki,yadegari2015generic,salwan2018symbolic,blazytko2017syntia,ding2019asm2vec,yang2021asteria,Wang_2025,sudhir2026pushantracefreedeobfuscationvirtualizationobfuscated,altamura2025tigress}.

Meanwhile, the rise of Apple Silicon, ARM-based servers, including the growing numbers of embedded architectures (e.g. RISC-V) has expanded the scope of binary analysis beyond x86 and Linux~\cite{riscvsv,waterman2011risc,apple-m5}. A fully comprehensive cross-build dataset would extend along the platform axis as well, covering macOS and ARM/RISC-V in addition to the configurations we currently provide.

\section{Acknowledgments}

This work was supported by NSF award CCF-2316159. This work was also supported in part through computational resources provided by Syracuse University. The authors gratefully acknowledge use of the OrangeGrid / HTC Campus Grid, supported by NSF award ACI-1341006, and technical support from Syracuse University's Cyberinfrastructure Engineer, supported by NSF award ACI-1541396.

\printbibliography

\newpage
\appendix

\input{appendix}

\end{document}

%% file: style_bibtex.tex
\usepackage[natbib=true, style=numeric, doi=false, isbn=false, url=true, eprint=false,maxnames=99,maxbibnames=99, sortcites=true, ]{biblatex}

\AtEveryBibitem{
  \clearfield{pages}
  \clearfield{volume}
  \clearfield{number}
  \clearfield{month}
  \clearfield{series}
  \clearlist{publisher}
  \clearlist{organization}
  \clearlist{address}
  \clearname{editor}
  \clearlist{location}
}

%% file: appendix.tex
\section{Technical appendices}

\begin{table}[H]
\centering
\footnotesize
\caption{All CVE IDs in paper.}\label{tab:cve-list-eval23}
\begin{tabular}{@{}l @{\quad} l @{\quad} l @{\quad} l @{\quad} l@{}}
\toprule
CVE-2013-0340 & CVE-2021-33468 & CVE-2023-25435 & CVE-2024-32619 & CVE-2025-3160 \\
CVE-2014-2497 & CVE-2021-3598 & CVE-2023-26965 & CVE-2024-32620 & CVE-2025-3196 \\
CVE-2016-3751 & CVE-2021-3605 & CVE-2023-26966 & CVE-2024-32621 & CVE-2025-3277 \\
CVE-2016-5767 & CVE-2021-38115 & CVE-2023-27102 & CVE-2024-32622 & CVE-2025-3549 \\
CVE-2017-12652 & CVE-2021-3933 & CVE-2023-27103 & CVE-2024-32623 & CVE-2025-43967 \\
CVE-2017-17506 & CVE-2021-40145 & CVE-2023-2731 & CVE-2024-32624 & CVE-2025-48072 \\
CVE-2017-6363 & CVE-2021-40529 & CVE-2023-2908 & CVE-2024-33873 & CVE-2025-48073 \\
CVE-2017-7890 & CVE-2021-40530 & CVE-2023-29416 & CVE-2024-33874 & CVE-2025-48074 \\
CVE-2017-9233 & CVE-2021-40812 & CVE-2023-29417 & CVE-2024-33875 & CVE-2025-48174 \\
CVE-2018-1000222 & CVE-2021-43519 & CVE-2023-29419 & CVE-2024-33876 & CVE-2025-48175 \\
CVE-2018-1311 & CVE-2021-43666 & CVE-2023-29420 & CVE-2024-33877 & CVE-2025-50952 \\
CVE-2018-14553 & CVE-2021-44647 & CVE-2023-29421 & CVE-2024-34250 & CVE-2025-5165 \\
CVE-2018-17233 & CVE-2021-44964 & CVE-2023-30774 & CVE-2024-34251 & CVE-2025-5166 \\
CVE-2018-17234 & CVE-2021-45960 & CVE-2023-31038 & CVE-2024-34402 & CVE-2025-5167 \\
CVE-2018-17237 & CVE-2021-45985 & CVE-2023-31974 & CVE-2024-34403 & CVE-2025-5168 \\
CVE-2018-17432 & CVE-2021-46141 & CVE-2023-35784 & CVE-2024-34703 & CVE-2025-5200 \\
CVE-2018-17435 & CVE-2021-46142 & CVE-2023-35789 & CVE-2024-40724 & CVE-2025-5202 \\
CVE-2018-17437 & CVE-2021-46143 & CVE-2023-39327 & CVE-2024-41672 & CVE-2025-5204 \\
CVE-2018-17438 & CVE-2021-46822 & CVE-2023-39329 & CVE-2024-45490 & CVE-2025-54812 \\
CVE-2018-25032 & CVE-2021-46880 & CVE-2023-43887 & CVE-2024-45491 & CVE-2025-54813 \\
CVE-2018-5711 & CVE-2022-0561 & CVE-2023-47466 & CVE-2024-45492 & CVE-2025-54874 \\
CVE-2019-11038 & CVE-2022-0562 & CVE-2023-47471 & CVE-2024-45679 & CVE-2025-56226 \\
CVE-2019-15903 & CVE-2022-0865 & CVE-2023-48105 & CVE-2024-48423 & CVE-2025-59375 \\
CVE-2019-18609 & CVE-2022-0908 & CVE-2023-4863 & CVE-2024-48424 & CVE-2025-61143 \\
CVE-2019-20454 & CVE-2022-0909 & CVE-2023-50980 & CVE-2024-48425 & CVE-2025-61144 \\
CVE-2019-25048 & CVE-2022-0924 & CVE-2023-51792 & CVE-2024-48426 & CVE-2025-61147 \\
CVE-2019-25049 & CVE-2022-1586 & CVE-2023-52284 & CVE-2024-50382 & CVE-2025-6120 \\
CVE-2019-6706 & CVE-2022-1587 & CVE-2023-52425 & CVE-2024-50383 & CVE-2025-62408 \\
CVE-2019-6977 & CVE-2022-1622 & CVE-2023-52426 & CVE-2024-50602 & CVE-2025-62600 \\
CVE-2019-6978 & CVE-2022-2056 & CVE-2023-53154 & CVE-2024-50612 & CVE-2025-6269 \\
CVE-2019-7317 & CVE-2022-22822 & CVE-2023-5841 & CVE-2024-5171 & CVE-2025-64505 \\
CVE-2019-8396 & CVE-2022-22823 & CVE-2023-6992 & CVE-2024-53425 & CVE-2025-64506 \\
CVE-2020-10810 & CVE-2022-22824 & CVE-2023-7256 & CVE-2024-56827 & CVE-2025-64720 \\
CVE-2020-10811 & CVE-2022-22825 & CVE-2024-1580 & CVE-2024-7006 & CVE-2025-65018 \\
CVE-2020-12762 & CVE-2022-22826 & CVE-2024-24806 & CVE-2024-8006 & CVE-2025-6516 \\
CVE-2020-14409 & CVE-2022-22827 & CVE-2024-25269 & CVE-2025-0838 & CVE-2025-66293 \\
CVE-2020-14410 & CVE-2022-22844 & CVE-2024-25431 & CVE-2025-11961 & CVE-2025-67899 \\
CVE-2020-15166 & CVE-2022-23852 & CVE-2024-25629 & CVE-2025-11964 & CVE-2025-68431 \\
CVE-2020-15389 & CVE-2022-23990 & CVE-2024-27532 & CVE-2025-12495 & CVE-2025-6965 \\
CVE-2020-15945 & CVE-2022-25235 & CVE-2024-28182 & CVE-2025-12839 & CVE-2025-8835 \\
CVE-2020-24342 & CVE-2022-25236 & CVE-2024-28231 & CVE-2025-12840 & CVE-2025-8836 \\
CVE-2020-27814 & CVE-2022-25308 & CVE-2024-28757 & CVE-2025-2151 & CVE-2025-8837 \\
CVE-2020-27823 & CVE-2022-25309 & CVE-2024-29157 & CVE-2025-2152 & CVE-2026-22801 \\
CVE-2020-27824 & CVE-2022-25310 & CVE-2024-29158 & CVE-2025-2591 & CVE-2026-25646 \\
CVE-2020-27841 & CVE-2022-25313 & CVE-2024-29159 & CVE-2025-2592 & CVE-2026-25835 \\
CVE-2020-27842 & CVE-2022-25314 & CVE-2024-29161 & CVE-2025-27091 & CVE-2026-26200 \\
CVE-2020-27843 & CVE-2022-25315 & CVE-2024-29162 & CVE-2025-2750 & CVE-2026-26981 \\
CVE-2020-27844 & CVE-2022-25761 & CVE-2024-29164 & CVE-2025-2751 & CVE-2026-27135 \\
CVE-2020-27845 & CVE-2022-28506 & CVE-2024-29166 & CVE-2025-2754 & CVE-2026-27171 \\
CVE-2020-36400 & CVE-2022-28805 & CVE-2024-31047 & CVE-2025-2756 & CVE-2026-27489 \\
CVE-2020-6851 & CVE-2022-3171 & CVE-2024-31744 & CVE-2025-2757 & CVE-2026-27622 \\
CVE-2020-8112 & CVE-2022-33099 & CVE-2024-3203 & CVE-2025-28162 & CVE-2026-29022 \\
CVE-2021-20234 & CVE-2022-3510 & CVE-2024-3204 & CVE-2025-28164 & CVE-2026-32877 \\
CVE-2021-20235 & CVE-2022-35737 & CVE-2024-32605 & CVE-2025-29087 & CVE-2026-32884 \\
CVE-2021-20237 & CVE-2022-3597 & CVE-2024-32607 & CVE-2025-2913 & CVE-2026-33165 \\
CVE-2021-20298 & CVE-2022-3627 & CVE-2024-32608 & CVE-2025-2914 & CVE-2026-33416 \\
CVE-2021-20304 & CVE-2022-37434 & CVE-2024-32609 & CVE-2025-2915 & CVE-2026-33636 \\
CVE-2021-29338 & CVE-2022-3970 & CVE-2024-32610 & CVE-2025-2923 & CVE-2026-34379 \\
CVE-2021-30473 & CVE-2022-40090 & CVE-2024-32611 & CVE-2025-2924 & CVE-2026-34380 \\
CVE-2021-30475 & CVE-2022-40674 & CVE-2024-32612 & CVE-2025-2926 & CVE-2026-34543 \\
CVE-2021-3246 & CVE-2022-43680 & CVE-2024-32613 & CVE-2025-3015 & CVE-2026-34544 \\
CVE-2021-32765 & CVE-2022-44370 & CVE-2024-32614 & CVE-2025-3016 & CVE-2026-34588 \\
CVE-2021-33454 & CVE-2022-48437 & CVE-2024-32615 & CVE-2025-31115 & CVE-2026-34589 \\
CVE-2021-33455 & CVE-2023-1999 & CVE-2024-32616 & CVE-2025-31498 & CVE-2026-34876 \\
CVE-2021-33461 & CVE-2023-25433 & CVE-2024-32617 & CVE-2025-3158 & CVE-2026-34877 \\
CVE-2021-33463 & CVE-2023-25434 & CVE-2024-32618 & CVE-2025-3159 &  \\
\bottomrule
\end{tabular}
\end{table}

\begin{table}[H]
\centering
\footnotesize
\caption{Manually inspected CVE IDs list.}
\label{tab:cve100_inspected}
\begin{tabular}{lllll}
\toprule
CVE-2016-5767 & CVE-2017-9233 & CVE-2018-17233 & CVE-2018-17234 & CVE-2018-17237 \\
CVE-2018-17432 & CVE-2018-17435 & CVE-2019-15903 & CVE-2019-18609 & CVE-2020-14409 \\
CVE-2020-14410 & CVE-2020-24342 & CVE-2020-27841 & CVE-2020-27844 & CVE-2020-36400 \\
CVE-2020-8112 & CVE-2021-20234 & CVE-2021-20237 & CVE-2021-20304 & CVE-2021-29338 \\
CVE-2021-30473 & CVE-2021-33455 & CVE-2021-3605 & CVE-2021-38115 & CVE-2021-40145 \\
CVE-2021-45960 & CVE-2021-45985 & CVE-2021-46141 & CVE-2021-46143 & CVE-2022-0562 \\
CVE-2022-22824 & CVE-2022-22827 & CVE-2022-22844 & CVE-2022-25308 & CVE-2022-25313 \\
CVE-2022-25314 & CVE-2022-25315 & CVE-2022-28506 & CVE-2022-3171 & CVE-2022-35737 \\
CVE-2022-40674 & CVE-2023-27102 & CVE-2023-27103 & CVE-2023-29417 & CVE-2023-29420 \\
CVE-2023-31038 & CVE-2023-35789 & CVE-2023-39329 & CVE-2023-6992 & CVE-2024-28231 \\
CVE-2024-29157 & CVE-2024-29158 & CVE-2024-29159 & CVE-2024-31047 & CVE-2024-3203 \\
CVE-2024-32605 & CVE-2024-32607 & CVE-2024-32610 & CVE-2024-32614 & CVE-2024-32615 \\
CVE-2024-32617 & CVE-2024-32619 & CVE-2024-33875 & CVE-2024-34251 & CVE-2024-40724 \\
CVE-2024-41672 & CVE-2024-45490 & CVE-2024-50382 & CVE-2024-53425 & CVE-2024-8006 \\
CVE-2025-11964 & CVE-2025-12495 & CVE-2025-12839 & CVE-2025-2914 & CVE-2025-2924 \\
CVE-2025-2926 & CVE-2025-3015 & CVE-2025-3196 & CVE-2025-43967 & CVE-2026-34876 \\
CVE-2025-50952 & & & &\\
\bottomrule
\end{tabular}
\end{table}

\begin{table}[H]
\centering
\small
\caption{Total packages in the DeepHistory dataset.}
\label{tab:packages}
\begin{tabular}{llll}
\toprule
\texttt{7bitdi} & \texttt{flatbuffers} & \texttt{librdkafka} & \texttt{poly2tri} \\
\texttt{abseil} & \texttt{flecs} & \texttt{libressl} & \texttt{proj} \\
\texttt{ada} & \texttt{fmt} & \texttt{libsigcpp} & \texttt{protobuf} \\
\texttt{ade} & \texttt{foonathan-memory} & \texttt{libsndfile} & \texttt{pthreadpool} \\
\texttt{aeron} & \texttt{freetype} & \texttt{libsodium} & \texttt{pugixml} \\
\texttt{ags} & \texttt{fribidi} & \texttt{libsrtp} & \texttt{pystring} \\
\texttt{alembic} & \texttt{fruit} & \texttt{libsvm} & \texttt{qcbor} \\
\texttt{amqp-cpp} & \texttt{ftxui} & \texttt{libsvtav1} & \texttt{qhull} \\
\texttt{antlr4-cppruntime} & \texttt{g3log} & \texttt{libtiff} & \texttt{qr-code-generator} \\
\texttt{apr} & \texttt{geographiclib} & \texttt{libusb} & \texttt{quirc} \\
\texttt{apr-util} & \texttt{geos} & \texttt{libuv} & \texttt{rabbitmq-c} \\
\texttt{argtable3} & \texttt{gflags} & \texttt{libwebp} & \texttt{rapidyaml} \\
\texttt{armadillo} & \texttt{giflib} & \texttt{libxlsxwriter} & \texttt{raylib} \\
\texttt{arsenalgear} & \texttt{ginkgo} & \texttt{libyuv} & \texttt{re2} \\
\texttt{asmjit} & \texttt{glad} & \texttt{libzen} & \texttt{redis-plus-plus} \\
\texttt{assimp} & \texttt{glfw} & \texttt{libzip} & \texttt{rmlui} \\
\texttt{base64} & \texttt{glog} & \texttt{libzippp} & \texttt{roaring} \\
\texttt{bdwgc} & \texttt{gm2calc} & \texttt{lief} & \texttt{rocksdb} \\
\texttt{blend2d} & \texttt{gtest} & \texttt{lightgbm} & \texttt{rotor} \\
\texttt{boost} & \texttt{h3} & \texttt{log4cplus} & \texttt{sail} \\
\texttt{botan} & \texttt{hazelcast-cpp-client} & \texttt{log4cxx} & \texttt{screen\_capture\_lite} \\
\texttt{box2d} & \texttt{hdf5} & \texttt{lua} & \texttt{sdl} \\
\texttt{brotli} & \texttt{hdrhistogram-c} & \texttt{lunasvg} & \texttt{sdl\_ttf} \\
\texttt{bzip2} & \texttt{hidapi} & \texttt{lz4} & \texttt{sfml} \\
\texttt{bzip3} & \texttt{highs} & \texttt{matio} & \texttt{simdjson} \\
\texttt{c-ares} & \texttt{highway} & \texttt{mbedtls} & \texttt{simdutf} \\
\texttt{c-blosc} & \texttt{hiredis} & \texttt{md4c} & \texttt{snappy} \\
\texttt{c-blosc2} & \texttt{http\_parser} & \texttt{meshoptimizer} & \texttt{sobjectizer} \\
\texttt{c4core} & \texttt{imath} & \texttt{mimalloc} & \texttt{spdlog} \\
\texttt{calceph} & \texttt{imgui} & \texttt{miniz} & \texttt{spirv-cross} \\
\texttt{cargs} & \texttt{implot} & \texttt{minizip} & \texttt{spirv-tools} \\
\texttt{catch2} & \texttt{ixwebsocket} & \texttt{mozjpeg} & \texttt{sqlite3} \\
\texttt{ceres-solver} & \texttt{jasper} & \texttt{mpdecimal} & \texttt{strawberryperl} \\
\texttt{cfitsio} & \texttt{jbig} & \texttt{msgpack-c} & \texttt{taglib} \\
\texttt{cgltf} & \texttt{json-c} & \texttt{msys2} & \texttt{tcl} \\
\texttt{charls} & \texttt{json-schema-validator} & \texttt{nasm} & \texttt{tinyspline} \\
\texttt{cjson} & \texttt{jsoncpp} & \texttt{nghttp3} & \texttt{tinyxml2} \\
\texttt{clipper} & \texttt{kuba-zip} & \texttt{ninja} & \texttt{tracy} \\
\texttt{clipper2} & \texttt{leptonica} & \texttt{nlopt} & \texttt{tree-sitter} \\
\texttt{cpp-optparse} & \texttt{lerc} & \texttt{nng} & \texttt{tree-sitter-c} \\
\texttt{cppcommon} & \texttt{lexbor} & \texttt{nsync} & \texttt{uriparser} \\
\texttt{cpptrace} & \texttt{libaesgm} & \texttt{octomap} & \texttt{utf8proc} \\
\texttt{cpu\_features} & \texttt{libaom-av1} & \texttt{ogg} & \texttt{vorbis} \\
\texttt{cpuinfo} & \texttt{libassert} & \texttt{onnx} & \texttt{vsg} \\
\texttt{cryptopp} & \texttt{libatomic\_ops} & \texttt{open62541} & \texttt{vulkan-loader} \\
\texttt{cwalk} & \texttt{libavif} & \texttt{openal} & \texttt{vvenc} \\
\texttt{dataframe} & \texttt{libde265} & \texttt{openal-soft} & \texttt{wasm-micro-runtime} \\
\texttt{dav1d} & \texttt{libdeflate} & \texttt{openblas} & \texttt{wasmtime} \\
\texttt{djinni-support-lib} & \texttt{libdwarf} & \texttt{opencl-icd-loader} & \texttt{winflexbison} \\
\texttt{double-conversion} & \texttt{libfdk\_aac} & \texttt{opencolorio} & \texttt{xerces-c} \\
\texttt{draco} & \texttt{libgd} & \texttt{openddl-parser} & \texttt{xz\_utils} \\
\texttt{drwav} & \texttt{libheif} & \texttt{openexr} & \texttt{yaml-cpp} \\
\texttt{duckdb} & \texttt{libjpeg} & \texttt{openh264} & \texttt{yasm} \\
\texttt{eastl} & \texttt{libjpeg-turbo} & \texttt{openjpeg} & \texttt{yyjson} \\
\texttt{embree3} & \texttt{libjxl} & \texttt{openmesh} & \texttt{z3} \\
\texttt{expat} & \texttt{libmaxminddb} & \texttt{opus} & \texttt{zeromq} \\
\texttt{ezc3d} & \texttt{libmp3lame} & \texttt{pcre} & \texttt{zimg} \\
\texttt{fast-cdr} & \texttt{libnghttp2} & \texttt{pcre2} & \texttt{zlib} \\
\texttt{fast-dds} & \texttt{libpcap} & \texttt{pdf-writer} & \texttt{zlib-ng} \\
\texttt{fastgltf} & \texttt{libpng} & \texttt{perfetto} & \texttt{zstd} \\
\texttt{fftw} & \texttt{libpq} & \texttt{pkgconf} & \texttt{zulu-openjdk} \\
\texttt{flac} & \texttt{libpqxx} & \texttt{plutovg} & \texttt{zxing-cpp} \\
\bottomrule
\end{tabular}
\end{table}

\section{Evaluation appendices}

\begin{table}[ht]
  \centering
  \label{tab:eval2-runtime-percentiles}
  \begin{tabular}{lrrr}
    \toprule
    Metric & $p_{50}$ & $p_{90}$ & $p_{99}$ \\
    \midrule
    Running time (s) & 77.3 & 573.5 & 1680.7 \\
    API call turns   & 26 & 105 & 232 \\
    \bottomrule
  \end{tabular}
  \caption{Eval~2 trial run time and agent turns on a random sample of 100 task.}
\end{table}

\begin{table}[H]
  \centering
  \small
  \begin{tabular}{lccc}
    \toprule
     & \texttt{ELF/clang} & \texttt{ELF/gcc} & \texttt{PE/msvc} \\
    \midrule
    \texttt{ELF/clang} & 0.53  & 0.45  & 0.29 \\
    \texttt{ELF/gcc} & 0.45  & 0.54  & 0.30 \\
    \texttt{PE/msvc} & 0.29  & 0.30  & 0.69 \\
    \bottomrule
  \end{tabular}
  \caption{Mean CodeBLEU between cross-build variants, aggregated by compiler family.}
  \label{tab:codebleu}
\end{table}

\begin{table}[H]
\centering
\setlength{\tabcolsep}{8pt}
\renewcommand{\arraystretch}{1.15}
\small
\setlength{\tabcolsep}{8pt}
\renewcommand{\arraystretch}{1.15}
\begin{tabular}{lrr}
\toprule
\textbf{Bucket} & \textbf{CVEs} & \textbf{Variants} \\
\midrule
Reference (Win + Linux) & 329 &    536 \\
\midrule
Diff\_Opt        & 209 &    621 \\
Diff\_Compiler   & 263 &    979 \\
Diff\_OS         & 173 &    932 \\
Diff\_Version    & 301 & 4{,}813 \\
Diff\_All        & 220 & 2{,}346 \\
\midrule
\textbf{Total}   & \textbf{329} & \textbf{10{,}227} \\
\bottomrule
\end{tabular}

\caption{Eval~3 cross-build variant counts.}
\label{tab:eval3-variants}
\end{table}

\begin{figure}[H]
    \centering
    \includegraphics[width=1\linewidth]{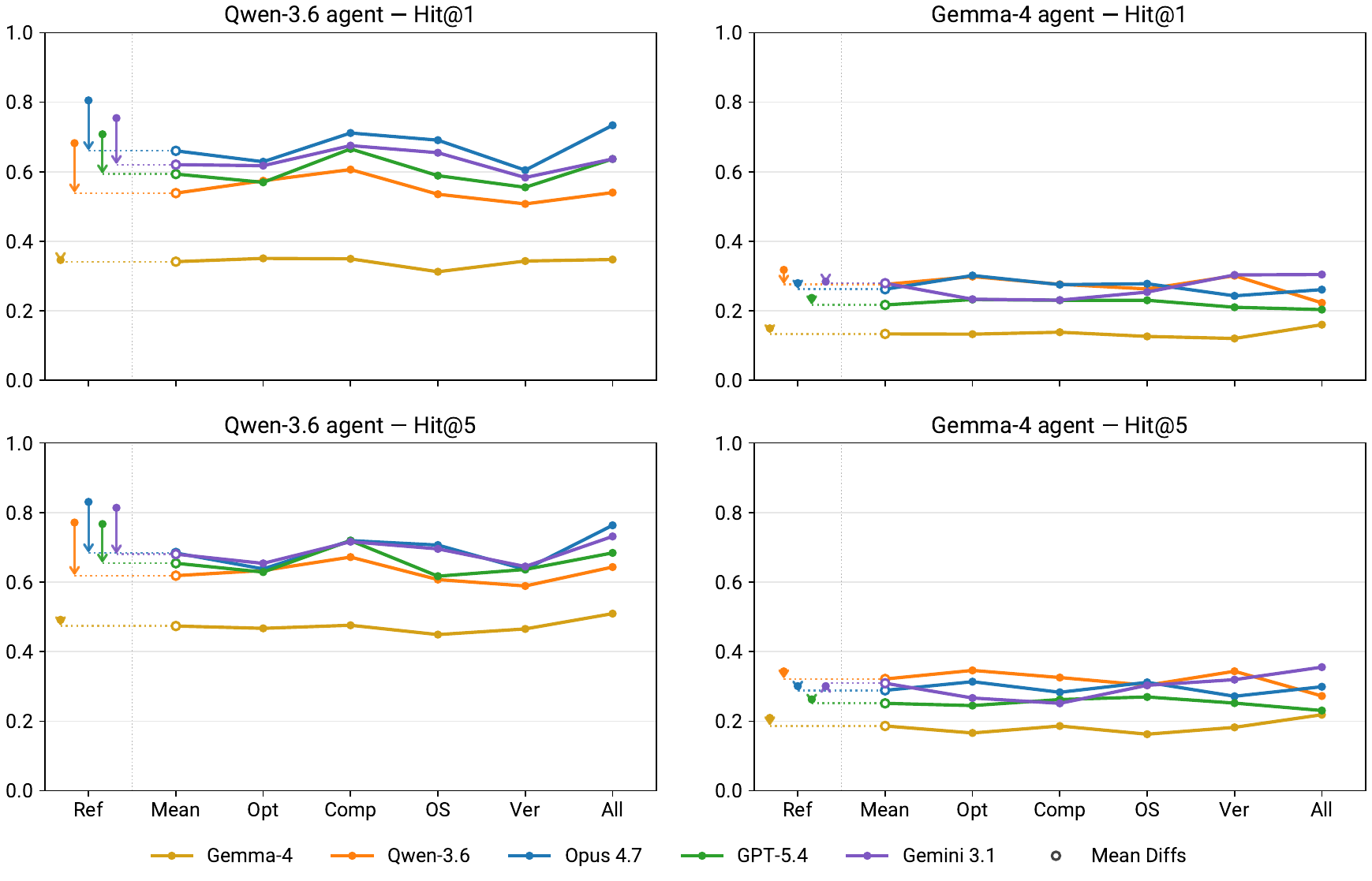}
    \caption{Eval 3 cross-build transfer comparison.}
\end{figure}

\section{Prompts}

\subsection{Strategy Generation Prompt}

\begin{lstlisting}[style=promptstyle]
You are a binary security researcher. Your task is to write a clear, step-by-step analysis strategy that a follower agent can execute to locate a known vulnerability in a stripped binary.

The follower agent has access to binary-analysis tools (decompiler, call graph, string and import enumeration, p-code) but no prior knowledge of this specific CVE. Your strategy describes a search procedure, not an answer - assume the agent does not know which function is vulnerable until your strategy guides them to it.

## Vulnerability Information

**CVE:** {cve_id}
**CWE:** {cwe_id} ({cwe_name})

**Description:**
{cve_description}

**Patch Diff:**
```diff
{patch_diff}
```

NOTE: The patch diff shows source-level function names like `{example_source_function}`. These names will NOT exist in stripped binaries - your strategy must describe the function by its behavior (imports it calls, strings it references, control flow), not its source name.

## Binary Analysis of Affected Function(s)

Two builds of the same vulnerable function are shown below. Your strategy must rely on properties stable across both - imported function names, string literals, call-graph relationships, semantic data and control-flow patterns.

### Reference build ({reference_build_key})
{reference_analysis}

### Cross-build variant ({variant_build_key})
{variant_analysis}

## Your Task

Write a strategy that a follower agent will execute against stripped binaries. The same vulnerability exists in every target, but compiled form - function boundaries, register allocation, inlining, control-flow shape - differs between builds. The strategy must identify the vulnerable function in *all* targets, including builds with different operating systems (Linux, Windows), compilers (GCC, Clang, MSVC), optimization levels (O0-O3), and software versions.

Some target binaries are PATCHED (vulnerability fixed). Your strategy must include a check that distinguishes vulnerable code from patched code, so the follower agent does not produce false positives on fixed binaries.

## Output Format

### Vulnerability Summary
One paragraph: what the vulnerability is and why it occurs.

### Search Strategy
Use as many steps as the vulnerability requires (typically 2-4). Each step should narrow the candidate pool.

**Step 1: [Name]**
- What to search for and why
- What results to expect

**Step 2: [Name]** (and so on)

### Verification Checklist
A list of yes/no facts the follower agent should confirm to distinguish a vulnerable candidate from a similar but patched function. e.g.,
- "Calls `memcpy` with size argument derived from attacker input:
  yes/no"
- "Has a bounds check before the dereference: yes/no"

### Fallback
If the primary approach fails (e.g. expected strings or imports are missing in a particular build), describe an alternative search strategy.

### Final Output
The follower agent must return exactly 5 candidate
functions, ranked by confidence. Describe how to produce that
many candidates - if the primary approach yields fewer, the
fallback or weaker signals should supply backup candidates.
\end{lstlisting}

\subsection{BinaryAPI Agent Documentation}

\begin{lstlisting}[style=promptstyle]
# BinaryAPI Reference

Read-only query interface for a stripped binary executable. All analysis is pre-computed; every method is a pure lookup.

Stripped binaries have auto-generated function names like `sub_401234`. Imported library symbols (e.g., `malloc`, `memcpy`) retain their original names.

## Construction

```python
api = BinaryAPI("/path/to/binary")
```

Accepts ELF or PE binaries. Analysis is cached by SHA-256 hash.

---

## Enumeration

### `list_functions() -> list[FunctionInfo]`

List all internal functions in the binary, sorted by address.

Each entry is a dict:
```python
{
    "name": "sub_4012a0",    # auto-generated identifier
    "address": "0x4012a0",   # hex entry point
    "size_bytes": 342,       # raw binary size
    "num_blocks": 12         # basic block count
}
```

### `get_imports() -> list[str]`

List all imported library function names, sorted alphabetically.

Example: `["free", "malloc", "memcpy", "printf", "strlen"]`

### `get_strings() -> list[str]`

List all string literals (>= 4 characters) found in the binary, sorted alphabetically.

---

## Property-Based Discovery

### `find_callers_of_import(import_name: str) -> list[str]`

Return internal functions that directly call the given imported function.

```python
api.find_callers_of_import("malloc")
# ["sub_401230", "sub_4015a0", "sub_401bc0"]
```

Returns `[]` if the import does not exist in this binary.

### `find_functions_referencing_string(s: str, case_sensitive: bool = False) -> list[str]`

Return functions that reference a string containing `s` (substring match). **Case-insensitive by default** - `s="XPM"` matches strings like `"xpm_load_image"`. Pass `case_sensitive=True` to require exact case (the literal bytes in the binary).

```python
api.find_functions_referencing_string("error")
# ["sub_401a00", "sub_402100"]

api.find_functions_referencing_string("XPM")               # -> finds xpm_*, XPM_*, Xpm_*, ...
api.find_functions_referencing_string("XPM", case_sensitive=True)  # -> only the exact 'XPM' bytes
```

Returns `[]` if no string contains the substring.

---

## Call-Graph Navigation

### `get_callees(func: str) -> list[str]`

Return all functions directly called by `func` (both imports and internal functions).

```python
api.get_callees("sub_401230")
# ["malloc", "memcpy", "sub_401100", "sub_401500"]
```

### `get_callers(func: str) -> list[str]`

Return all functions that contain a direct call to `func`.

```python
api.get_callers("sub_401230")
# ["sub_400f00", "sub_401800"]
```

---

## Code Inspection

### `decompile(func: str) -> str`

Return Ghidra-decompiled C pseudocode for `func`.

Variable names, expression order, and control-flow structure vary across compilers and optimization levels. Useful for human-readable understanding but less stable across builds than p-code.

```python
api.decompile("sub_401230")
# "void sub_401230(long param_1, int param_2) {\n  ..."
```

### `get_pcode(func: str) -> str`

Return high p-code (post-SSA intermediate representation) for `func`, grouped by basic block.

P-code uses:
- Sequential SSA variable names: `v0`, `v1`, `v2`, ...
- Architecture-independent operations: `LOAD`, `STORE`, `INT_ADD`, `INT_SUB`, `INT_MULT`, `INT_AND`, `INT_OR`, `INT_XOR`, `INT_LEFT`, `INT_RIGHT`, `INT_SRIGHT`, `INT_EQUAL`, `INT_NOTEQUAL`, `INT_LESS`, `INT_SLESS`, `INT_LESSEQUAL`, `INT_SLESSEQUAL`, `BOOL_NEGATE`, `BOOL_AND`, `BOOL_OR`, `FLOAT_ADD`, `FLOAT_SUB`, `FLOAT_MULT`, `FLOAT_DIV`, `CALL`, `CALLIND`, `CBRANCH`, `BRANCH`, `BRANCHIND`, `RETURN`, `COPY`, `CAST`, `SUBPIECE`, `INT_ZEXT`, `INT_SEXT`, `PIECE`, `PTRADD`, `PTRSUB`
- Labelled blocks: `blk_0`, `blk_1`, ... (consistent with `get_cfg`)
- Phi nodes at merge points: `PHI(v3, v7)`

**Annotation format** (every operand is a self-describing token):
- `:N` width suffix in bytes - `v0:4` is a 4-byte SSA value, `0x10:8` is an 8-byte constant.
- `@REG` / `@stack[off]` location suffix on SSA values that live in a non-unique address space - e.g. `v0:8@RCX` is the SSA value occupying the RCX register (typical Win64 first arg), `v3:8@stack[-0x10]` is a stack slot.  RAM-space globals skip the SSA name entirely and render as `0x{addr}:N` since the address is the identity.
- `<type>` suffix when the decompiler inferred a non-undefined data type - `v0:8<longlong *>` is an 8-byte SSA value typed as a pointer to longlong.
- Signedness is preserved through the opcode (`INT_DIV` vs `INT_SDIV`, `INT_LESS` vs `INT_SLESS`) and the condition operator (`<u` for unsigned, `<` for signed).

More stable across build variants than decompiled C, but watch for type/register noise - the SSA `vN` numbering and the operation skeleton are the most robust elements.

Example output:
```
blk_0:
    v1:1@ZF<bool> = INT_EQUAL v0:8@RCX<longlong>, 0:8
    CBRANCH blk_2, v0:8@RCX<longlong> == 0:8

blk_1:
    v2:8@RAX = PTRSUB 0:8, 0x18004a908:8
    BRANCH blk_3

blk_2:
    v3:8<longlong> = INT_ADD v0:8@RCX<longlong>, 24:8
    v4:8<longlong *> = CAST v3:8<longlong>
    v5:8<longlong> = LOAD [v4:8<longlong *>]

blk_3:
    v9:8@RAX = PHI(v2:8@RAX, v5:8<longlong>)
    RETURN v9:8@RAX
```

Reading the example: `v0:8@RCX<longlong>` is the first argument (Win64 calling convention puts arg0 in RCX), 8 bytes wide, typed `longlong`.  The PHI in `blk_3` merges two distinct SSA generations that both happen to live in RAX - the SSA names (`v2`, `v9`) keep the generations separate even though they share the register.

### `get_assembly(func: str) -> str`

Return raw disassembly (one instruction per line) for `func`. Architecture-native
mnemonics (x86/x64/ARM/...); operands use the native register set and
addressing syntax.

Use this when decompiled C hides architecturally-relevant detail: timing
side-channels, constant-time violations, calling-convention issues,
register clobbering, or compiler-emitted branch layout. For most bugs
decompiled C is sufficient - prefer `decompile` or `get_pcode` first.

```python
api.get_assembly("sub_401230")
# "0x401230  PUSH RBP\n0x401231  MOV RBP,RSP\n0x401234  SUB RSP,0x30\n..."
```

---

## Control-Flow Graph

### `get_cfg(func: str) -> CFGResult`

Return the intra-procedural control-flow graph of `func`.

```python
{
    "function": "sub_401230",
    "blocks": ["blk_0", "blk_1", "blk_2", "blk_3"],
    "entry_block": "blk_0",
    "edges": [
        {"source": "blk_0", "target": "blk_1", "edge_type": "branch_false"},
        {"source": "blk_0", "target": "blk_2", "edge_type": "branch_true"},
        {"source": "blk_1", "target": "blk_3", "edge_type": "fallthrough"},
        {"source": "blk_2", "target": "blk_3", "edge_type": "unconditional"}
    ]
}
```

Edge types: `fallthrough`, `branch_true`, `branch_false`, `unconditional`.

Block labels match those in `get_pcode()` output.

---

## Regex Search

### `search_decompiled(pattern: str, limit: int = 200) -> SearchResults`

Apply a Python regex `pattern` line-by-line across all decompiled output. Returns matches grouped by function, capped at `limit` total matches.

```python
api.search_decompiled(r"memcpy\(.*,.*,.*\)")
# {
#   "results": [
#     {
#       "function": "sub_401230",
#       "match_count": 2,
#       "matches": [
#         {"function": "sub_401230", "line_number": 15,
#          "line_content": "  memcpy(local_buf, param_1, param_2);",
#          "match_text": "memcpy(local_buf, param_1, param_2)"},
#         ...
#       ]
#     }
#   ],
#   "total_match_count": 2,
#   "truncated": false,
#   "limit": 200
# }
```

When `truncated` is `true`, the search stopped at `limit` matches and there may be more. Refine the pattern (more specific regex) or raise `limit` on the next call. Note that `match_count` for the last function in `results` may understate its true count when truncation hit mid-function.

### `search_pcode(pattern: str, limit: int = 200) -> SearchResults`

Apply a Python regex `pattern` line-by-line across all p-code output. Same return shape and truncation contract as `search_decompiled`.

```python
api.search_pcode(r"CALL memcpy")
# Functions that call memcpy, with exact p-code lines
```

### `search_assembly(pattern: str, limit: int = 200) -> SearchResults`

Apply a Python regex `pattern` line-by-line across all disassembly. Same return shape and truncation contract as `search_decompiled`.

```python
api.search_assembly(r"^[^;]*\bRDTSC\b")
# Functions that contain RDTSC (time-stamp counter reads)
```

---

## Error Handling

| Scenario | Exception |
|----------|-----------|
| Function name not found | `KeyError` |
| Method called on an imported function (e.g., `decompile("malloc")`) | `KeyError` |
| Invalid regex pattern | `ValueError` |
| Import not in binary (`find_callers_of_import`) | Returns `[]` |
| No string match (`find_functions_referencing_string`) | Returns `[]` |

No other exceptions during normal query operation. I/O and analysis errors surface only at construction time.

---

## What Survives Stripping

| Property | Survives? | Stability Across Builds |
|----------|-----------|------------------------|
| Imported function names | Yes | High |
| String literals | Yes | High |
| Call graph structure | Yes | Moderate |
| P-code operations | Yes | High (arch-independent) |
| CFG structure | Yes | Moderate |
| Internal function names | **No** (auto-generated) | None |
| Addresses/offsets | Yes but **change every build** | None |
| Decompiled C variable names | Yes but **vary by compiler** | Low |

\end{lstlisting}

\subsection{Agent Solo Mode Prompt}

\begin{lstlisting}[style=promptstyle]

You are a binary security researcher. Your task is to locate a
security vulnerability in a stripped binary by querying it through
an analysis API. You are given no prior knowledge about the
vulnerability - you must discover it on your own.

## Important Context

The target binary is STRIPPED - internal function names are
auto-generated identifiers like "sub_401234". You must locate
suspicious functions by querying the binary's properties:
imports, strings, call graph, decompiled code, p-code, and
raw disassembly.

## Available API

You have access to the following API. Call any method at any time.
There is no limit on the number of calls you can make, and no cap
on the number of turns. A single run is bounded only by a 1 hour
wall-clock budget.

{api_documentation}

### Paging

`list_functions`, `get_imports`, and `get_strings` are paginated
via `offset` and `limit` (default limit 100). Each response tells
you the `total`, the current `offset`, how many items were
`returned`, and the `next_offset` to fetch the following page
(or null when exhausted). Walk through all pages if you need a
global view.

### Managing context

Every tool result you receive is prefixed with `[tag=tN]`. If a
result has served its purpose (e.g., a large decompilation you
have already digested), call `discard_tool_result(tag="tN")` to
replace it with a short placeholder and free context for later
queries. Only discard what you no longer need - discarded results
cannot be recovered in the same run.

## Instructions

Analyze the binary step by step:
1. Start by exploring the binary (imports, strings, functions).
   Use paging and `discard_tool_result` freely - prefer many
   targeted queries over hoarding large raw dumps.
2. Look for risky patterns: unchecked sizes passed to memcpy /
   strcpy, use-after-free, integer overflow before allocation,
   format strings, missing bounds checks, etc.
3. Narrow down candidates through API queries.
4. Inspect promising functions via decompile(), get_pcode(), or
   get_assembly().
5. When confident, provide your final answer.

After each API call, you will see the results. Decide your next
action based on what you learn.

When you are ready, output your final answer as:

CANDIDATES: [func1, func2, ...]

Return exactly 5 candidates, ranked from most likely
to least likely. If you have fewer than 5 strong
candidates, fill remaining slots with your best guesses.
\end{lstlisting}

\subsection{Agent Strategy Guided Mode Prompt}

\begin{lstlisting}[style=promptstyle]
You are a binary analysis assistant. A senior security researcher
has written a strategy for locating a known vulnerability in a
stripped binary. Your job is to follow this strategy step by step,
making the appropriate API calls and reporting results.

## Strategy

{strategy_document}

## Important Context

The target binary is STRIPPED - internal function names are
auto-generated identifiers like "sub_401234". The original source
function names referenced in the strategy do not exist in the
binary. When the strategy says "find functions that do X", you
must translate that into appropriate API calls.

## Available API

{api_documentation}

### Paging

`list_functions`, `get_imports`, and `get_strings` are paginated
via `offset` and `limit` (default 100). Each response includes
`total`, `offset`, `returned`, and `next_offset`.

### Managing context

Every tool result is prefixed with `[tag=tN]`. Call
`discard_tool_result(tag="tN")` to drop a result you no longer
need. The run has no turn cap and no tool-result size cap - only
a 1 hour wall-clock budget.

## Instructions

Follow the strategy step by step:
1. Read each step of the strategy carefully
2. Translate it into one or more API calls
3. Report the results of each call
4. If a step returns empty results, check if the strategy
   provides a fallback. If so, follow the fallback. If not,
   note the failure and proceed to the next step.
5. After completing all steps, provide your final answer

Do NOT add your own vulnerability analysis beyond what the
strategy describes. Your role is to execute the strategy
faithfully, not to independently reason about the vulnerability.

When you are ready, output your final answer as:

CANDIDATES: [func1, func2, ...]

Return exactly 5 candidates, ranked from most likely
to least likely. If you have fewer than 5 strong
candidates, fill remaining slots with your best guesses.
\end{lstlisting}